# CHAMPION: Chalmers Hierarchical Atomic, Molecular, Polymeric & Ionic Analysis Toolkit

Rasmus Andersson[a,b], Fabian Årén[a,b], Alejandro A. Franco[c,d,e,f], and Patrik Johansson[a,b,e]

[a]*Department of Physics, Chalmers University of Technology, SE-412 96 Gothenburg, Sweden*
[b]*Compular AB, SE-416 56 Gothenburg, Sweden*
[c]*Laboratoire de Réactivité et Chimie des Solides (LRCS), Université de Picardie Jules Verne UMR CNRS 7314, Hub de l'Energie, 15 Rue Baudelocque, 80039 Amiens, France*
[d]*Réseau sur le Stockage Electrochimique de l'Energie (RS2E), FR CNRS 3459, Hub de l'Energie, 15 Rue Baudelocque, 80039 Amiens, France*
[e]*Alistore-ERI European Research Institute, CNRS FR 3104, Hub de l'Energie, 15 Rue Baudelocque, 80039 Amiens, France*
[f]*Institut Universitaire de France, 103 Boulevard Saint-Michel, 75005 Paris, France*

We present CHAMPION: a software developed to automatically detect time-dependent bonds between atoms based on their dynamics, classify the local graph topology around them, and analyze the physicochemical properties of these topologies by statistical physics. In stark contrast to methodologies where bonds are detected based on static conditions such as cut-off distances, CHAMPION considers pairs of atoms to be bound only if they move together and act as a bound pair over time. Furthermore, the time-dependent global bond graph is possible to split into dynamically shifting connected components or subgraphs around a certain chemical motif and thereby allow the physicochemical properties of each such topology to be analyzed by statistical physics. Applicable to condensed matter and liquids in general, and electrolytes in particular, this allows both quantitative and qualitative descriptions of local structure, as well as dynamical processes such as speciation and diffusion. We present here a detailed overview of CHAMPION, including its underlying methodology, implementation and capabilities.

## 1. Introduction

Many scientifically and technologically important materials and liquids today are complex in terms of their intermolecular structure and dynamics. Examples include electrolytes in electrochemical devices, liquid solutions, dispersions and emulsions, drug delivery systems, and most polymeric systems including gels, plastics, elastomers and fibres, and even liquid water when the dynamic network of hydrogen bonds is taken into account.

One particular example, which also motivated the development of the software reported here (CHAMPION: Chalmers Hierarchical Atomic, Molecular, Polymeric & Ionic Anaylsis Toolkit) and which will be used as example throughout, is electrolytes for lithium-ion batteries (LIB) as well as electrolytes for various next generation batteries.[1] To enable the reader to fully understand the need for CHAMPION, and what it must be able to do – we here describe these systems to some detailed extent before progressing to the software itself. The purpose of the electrolyte in a battery is basically to enable transport of ions between the electrodes. In addition, many other requirements must be met: wide liquidus range, electrochemical, chemical, as well as thermal stability, low toxicity and environmental impact, low vapour pressure and flammability, etc.

The current state-of-the-art LIB electrolyte is based on 1 M LiPF$_6$ in a solvent mixture of cyclic and linear carbonates, the typical and original examples thereof being ethylene carbonate (EC) and dimethyl carbonate (DMC) (Fig. 1),[1]

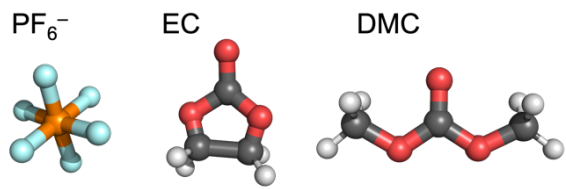

*Figure 1. Common constituents of LIB electrolytes.*

and several additives that together form a functional electrolyte.[2] This relatively simple, in terms of main composition, electrolyte also has the advantage that it is relatively easy to understand and model at the molecular level: the Li$^+$ ions tend to be fourfold coordinated by solvent molecules forming its first solvation shell[3] and the cation transport more or less follows the Stokes-Einstein relation,[4] where the first solvation shell is transported at a rate limited by the electrolyte viscosity and the first solvation shell radius. Some Li$^+$ ions coordinate to a single anion in the place of one of the solvent molecules, *i.e.* form ion-pairs, but also a few larger aggregates form.[3,1] Most of this speciation and coordination, in the bulk, is relatively stable and long-lived, and thus does not necessarily complicate the overall picture of the electrolyte structure or ion transport. The ion conductivity has a maximum at ca. 1 M salt concentration, mainly as this renders both many fully solvent solvated Li$^+$ ions and many uncoordinated solvent molecules that improve the fluidity, but this also has the drawback of the electrolyte being volatile and even

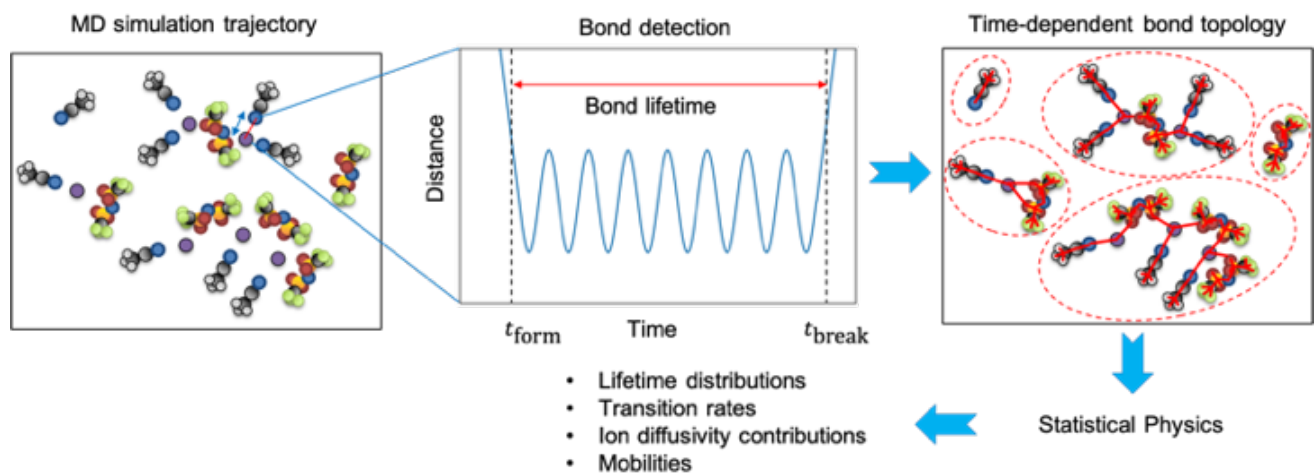

*Figure 2. Schematic of our novel method illustrated by lithium bis(trifluoromethane sulfonyl) imide (LiTFSI) in acetonitrile (ACN). Element colours: purple: Li, red: O, blue: N, grey: C, white: H, yellow: S, green: F. Reprinted with permission from the Journal of the Electrochemical Society.[11]*

flammable under abuse conditions.[5] Additionally, these electrolytes have limited electrochemical stability windows (ESWs) – especially in light of the high voltage electrodes under development to enable up to 5 V LIB cells.[6] It is, however, very difficult to modify the electrolyte composition substantially without reducing the overall performance.[1]

Recently, highly concentrated electrolytes (HCEs) have emerged as a possible route to improve both safety and performance; simply by increasing the salt concentration drastically, up to ca. 3-10 M, while often using the same electrolyte chemistry.[7,8] The change in concentration radically alters the local and time-dependent structure of the electrolyte resulting in different local structures with more ion-ion interactions, as there is less solvent available to coordinate Li$^+$, and also larger cohesive structures[9] – even percolating liquid networks.[10,11] This improves safety by having less volatile solvent present,[12,13] widens ESWs by altered electronic and interfacial structures,[14,15] but foremost also leads to completely different ion transport mechanisms; relying on structural reorganisation rather than viscosity-limited diffusion and electromigration.[11] While HCEs in general have both higher viscosities and lower total ionic conductivities than conventional electrolytes[16,17], the altered (ion) transport mechanisms may be more favourable to cations than anions, leading to overall acceptable performance.[18,19,17] To understand HCEs in their full complexity as well as other complex liquid battery electrolytes, such as ionic liquid (IL) based electrolytes,[20,21] solvate ionic liquids (SILs)[22], localized highly concentrated electrolytes (LHCEs),[23] and even gel and solid polymer electrolytes (GPEs and SPEs),[24,25,26] calls for new dynamic and atomically resolved tools. Similarly, rheological and other physicochemical properties of solutions, emulsions and dispersions can only be comprehensively understood with detailed knowledge of their disordered local structure and dynamics.[27] The same is true for dynamic processes such as dissolution in e.g. drug delivery systems[28] and to understand the resulting structure and mechanical properties of polymeric and other non-crystalline materials as functions of preparation conditions[29,30].

To address such questions through modelling, we use molecular dynamics (MD), either classical or *ab initio* (AIMD), to generate atomic trajectories, which we analyse by CHAMPION both structurally and dynamically by a combination of graph theory and statistical physics. In principle, any computational or experimental method that resolves the atomic configuration and dynamics on the relevant timescale(s) for the process(es) targeted can be used.

For LIB electrolytes statistical physics can reveal *e.g.* Li$^+$ ion diffusivity from mean squared displacements (MSDs).[31] While such computations are predictive, they do not provide any real understanding of the details of the ion transport mechanism – and especially not if the local structure changes at the same time-scale – as is the case for HCEs.[16,32,11] Hence, these local structures have to be discovered, *e.g.* by identifying which atoms are bound together. We here use an inclusive notion of bond including both covalent, electrostatic, or any other type of interaction that results in a pair of atoms moving together as a cohesive unit.

For a standard LIB electrolyte partial radial distribution functions (pRDFs) or other similar static distance criteria can be used to assess whether two atoms are bound.[33,34] However, this approach relies on prior knowledge of the speciation in terms of which types of atoms tend to form bonds and cannot treat cases where such knowledge is lacking.

To resolve this CHAMPION uses a unique dynamic bond detection method based on requiring bound atoms to oscillate about a well-defined equilibrium distance of one another for an extended period of time (Fig. 2). The bond is active for as long as this oscillatory motion persists. The dynamic nature of this method ensures that bonds are assigned only based on pairs of atoms which actually move together.

While many software suites exist for post-processing of MD trajectories[35,36,37,38] and are included in many MD

simulation software packages,[39,40,41] as well as some visualization software,[42] none of these, to the best of our knowledge, implement bond detection based on dynamics. From this unique dynamic detection of local structures several properties can be extracted (for a vast range of systems) by statistical physics, such as bond graph topologies and their populations, life-time distributions for topologies and bonds, transition rates, and, of special interest to LIB electrolytes, contributions to diffusive and structural ion transport and the effective diffusivity. Below we both outline the details of the methods and the theory behind CHAMPION, as well as illustrate the power of this analysis tool by applying it to the complex case of HCEs.

CHAMPION is written in high-performance C++20 and consists of a modular header-only library and a collection of text-configured command-line tools. The examples shown required between 10 min and 10 h computational time on a Macbook Pro from 2015 with 2.2 GHz Intel Core i7 and 16 GB memory. In all cases the analysis required considerably less computational resources than the corresponding MD simulations. CHAMPION is proprietary software owned by Compular Technologies AB, and not yet generally available to third party users, but we encourage interested readers to contact us to enquire about possible trials or collaborations.

## 2. Methods, Theory and Results

As outlined above CHAMPION is mainly geared towards the analysis of atomic trajectories, but it also includes a system builder for disordered condensed matter systems, to support the set-up of simulations. The simulations themselves, however, are currently outside the scope of CHAMPION, which instead is able to import trajectories in the .xyz format from any external simulation software.

Subsequently, the trajectories are post-processed in three consecutive steps: to assess the structural equilibration time, to detect time-dependent bonds, and to do the topological classification of structures.

After this post-processing, a number of different and mostly independent statistical physics-based analyses can be performed – herein we provide seven examples. Each provides best estimates and statistical uncertainties of the physicochemical properties targeted, including statistical inefficiency, but always assuming ergodicity – the reasonableness of this assumption has to be considered by the user. The full workflow for running a CHAMPION analysis is illustrated in Scheme 1.

Below the system builder, the three post-processing steps and the analysis parts of CHAMPION are explained; what each of them do and the main theory behind – including some limitations and caveats.

**System builder**

This part of CHAMPION basically is a pre-simulation utility for generating (random) starting geometries. The user first specifies the topologies and geometries of a set of molecules (or similar) with a specific stoichiometry and density. The user also controls the maximum allowed total

**CHAMPION Workflow**

1. Configure system builder (text file)
2. Run system builder → .xyz (1-10 min.)
3. Run simulations
4. Configure analyser (text file)
5. Run analyser: (10 min-10 h.)
   - Computes pRDFs
   - Computes structural equilibration time
   - Detects bonds
   - Builds global bond graph
   - Finds connected components
   - Finds neighbourhoods
   - Statistical physics analysis
- → analysis output (text file)

*Scheme 1. Full workflow for performing a CHAMPION analysis. All but the grey step are done using CHAMPION. Numbered steps are user initiated, bullet points are executed automatically by the software.*

number of atoms in the simulation cell and the CHAMPION algorithm creates as many stoichiometric units as allowed by this limit. The algorithm first places and orients all molecules by a uniform random distribution in a cubic periodic box with the side set to give the specified density. In order to avoid overlapping atoms, which may cause stability problems for the simulations, the positions and orientations of the molecules are subsequently relaxed by a gradient descent (conjugate gradient method) using a cost function designed to maximize atomic distances relative to atomic radii,

$$C_i = \sum_{i \neq j} H(R_i + R_j - d_{ij})\left(1 - \frac{d_{ij}}{R_i + R_j}\right)^2 \qquad (1)$$

where $H(x)$ is the Heaviside function, $R_i$ and $R_j$ are the van der Waals (vdW) radii of atoms $i$ and $j$ and $d_{ij}$ is the distance between atoms $i$ and $j$. The relaxation is performed iteratively by simultaneously translating and rotating one molecule at a time until all molecules have been relaxed a specified number of times. The resulting configuration output is available in .xyz and .pdb formats and can subsequently be imported into any simulation software supporting periodic cubic geometries.

**Post-processing step 1 – Structural equilibration time**

Since simulations most often start from a randomised geometry rather than a physically plausible one, the structure equilibration time must be assessed – and this might also be a property of interest by itself. CHAMPION assesses the equilibration via changes over time in the pRDFs

$$g_{ij}(r) = \frac{1}{n_0}\frac{n(r)}{4\pi r^2} \quad (2)$$

where $n(r)$ is the number density of neighbours of atom type $j$ on distance $r$ from atoms of type $i$ and the expression is normalised by the average bulk number density $n_0$ of species $j$.

The instantaneous structural error

$$\epsilon(t) = \sum_{i\neq j}\int_0^{r_{\text{cut-off}}}\left(g_{ij}(r,t)-\bar{g}_{ij}(r)\right)^2 dr \quad (3)$$

is defined as the sum of integrated squared deviations of the instantaneous pRDFs from their time averages. Assuming that our simulation is long enough, i.e. substantially longer than the equilibration time, these time averages are valid approximations to the equilibrated pRDFs. By curve-fitting $\log\epsilon(t)$ to a continuous piecewise linear curve with a negative slope before the cusp and a horizontal line past it, we obtain an estimate of the equilibration time as the point where the two lines intersect. In practice, in order to obtain a conservative starting point for data production, the equilibration time can be multiplied by a number slightly greater than 1 e.g. 1.2.

This scheme should work well to set a reasonable starting point for the production run when the trajectory is considerably longer than the structural equilibration time, structural degrees of freedom (DoFs) are at least as slow to equilibrate as any other DoFs of interest, and binary correlations equilibrate at the same rate as all relevant higher-order correlations.

If the first condition is not fulfilled, this can be seen in the overall appearance of the $\epsilon(t)$ curve. If it has not yet equilibrated it should be everywhere convex, with a minimum near the midpoint of the trajectory, rather than decreasing and thereafter stable. For the other requirements, the user needs to exercise their domain knowledge and judgement. There is always the possibility that fast DoF may have completely converged, while slower processes of equal or greater effect on the overall structure may not move perceptibly at all over the trajectory. In this case the equilibration curve will still look like the system has equilibrated fully.

**Post-processing step 2 – Time-dependent bond detection**

CHAMPION's main feature (pat. pend.[43]) is the ability to characterize the structure of any system topologically as a time-dependent set of bonds between atoms, which may form and break during the course of the trajectory. The algorithm works as follows: Two atoms of species $i$ and $j$, which are closer than the sum of a fraction (typically 0.8 in our analyses) of their vdW radii (Fig. 3a) over a particular period of time $[t, t+\tau)$ are considered bound if their mean distance during this time is within a tolerance of the first peak in the pRDF (Eq. 1) for species $i$ and $j$ (Fig. 3b)

$$(1-\alpha)r_{\text{peak}} \leq \frac{1}{\tau}\int_t^{t+\tau} d_{ij}(t)dt \leq (1+\alpha)r_{\text{peak}} \quad (4)$$

for tolerance α (typically a few percents of the half-width at half maximum (HWHM)).

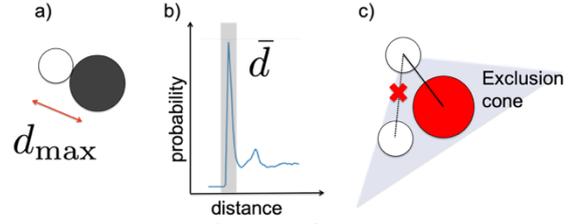

*Figure 3. CHAMPION's criteria for two atoms to be bound: a) the maximum distance over some duration is less than a cut-off value based on the species radii, b) the mean distance within this duration is within a tolerance of the first pRDF peak, and c) the bond is outside the exclusion cone formed about all sufficiently shorter bonds involving one of the atoms. The exclusion cone is determined by two parameters: an angle and a bond length ratio.*

In addition, a bond between atoms $i$ and $j$ is discarded if it is within a cone emanating from atom i with axis along a substantially shorter bond involving atom $i$ (the blocking bond), as that may indicate the cohesion to be caused by both atoms binding to a common neighbour rather than directly to each other, (Fig. 3c). How much shorter the blocking bond must be to eliminate the longer bond candidate, as well as the angle defining the exclusion cone, are user-set parameters. If all three conditions are fulfilled, there is a bond, and all that remains to be determined is the time of its birth and death. The starting guess is between the first entry and the final exit of $r_{\text{peak}}$ in $[t, t+\tau)$. This interval is subsequently expanded in both directions to the entry into/exit from the furthest turning distance between the starting guesses. This scheme for birth and death is designed to avoid spurious exits and re-entries to be registered by the algorithm.

A prerequisite for this algorithm to work is that the typical lifetime of a bond must stretch over many timesteps in order for the average distance to approach the pRDF peak. Indeed, the greater the average number of sampled timesteps for a bond, the tighter the bond length tolerance can be set.

This algorithm has been found to perform very well in detecting actual bonds and also in avoiding false positives, given an adequate choice of parameters that may have to be tailored to the system (mainly bond length tolerance, exclusion cone angle, and maximum allowed distance ratio with a shorter bond). A parameter debug switch can be activated to give detailed information on the grounds for accepting or rejecting each individual bond candidate. If necessary, pairs of elements that tend to register as bound, but that domain knowledge rejects as spurious, can be added to an exclusion list. This is, however, most often a symptom of ill-chosen parameters.

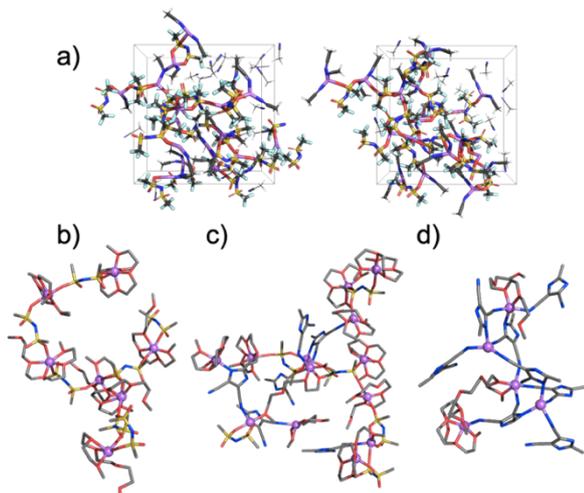

*Figure 4. Example structures (connected components) discovered by CHAMPION: a) Two snapshots from a simulation of LiTFSI in ACN at 1:2 salt:solvent molar fraction. Reprinted with permission from the Journal of the Electrochemical Society.[11] b)-d) Mixtures of LiTFSI and lithium 2-trifluoromethyl-4,5-dicyanoimidazole (LiTDI) in tetraglyme (G4) at 1:1 molar ratio, with LiTDI constituting b) 0%, c) 30%, and d) 100% of the salt. Reprinted with permission from Batteries & Supercaps.[10]*

**Post-processing step 3 – Structure classification**

After having established the set of bonds for each timestep, CHAMPION encodes this information into a time-dependent graph, henceforth referred to as the global bond graph of the system. All subsequent physicochemical characterisation is based on different ways of partitioning this graph into local subgraphs. All bond graphs, both global and local, are in CHAMPION represented as undirected graphs where the vertices represent atoms (and also encode their chemical species) and the edges represent bonds. Bond order is, however, not encoded by the edges, since it does not enter the bond detection algorithm.

CHAMPION supports two different kinds of decompositions of the global bond graph: components (Fig. 4) and local structures (Fig. 5). The former results from partitioning the graph in as many subgraphs as possible without cutting any edges, so that all vertices in each component are reachable from all other vertices in the component through one or more edges, but not from any other vertices in the graph. Local structures result from including all vertices and edges up to a certain graph distance (i.e. a certain maximum number of edges away) from a central topology. The preferred decomposition depends both on the nature of the system and the properties of interest – and is in the hand of the user.

CHAMPION also supports coarse graining of the global atomic bond graph into a molecular bond graph which analogously can be partitioned into components or local structures. The vertices of this graph are molecules, ions, monomers, or other user-provided topologies and the edges encode which of these topologies are directly bound. Here CHAMPION first identifies all subgraphs matching the user-provided topologies, then finds and classifies any unmatched components remaining after removing the matched topologies. This provides the basis for a semantic segmentation of the atomic bond graph, which is used to construct the molecular bond graph, which can of course also be mapped to the atomically resolved graph for further analysis or visualisation (Fig. 5).

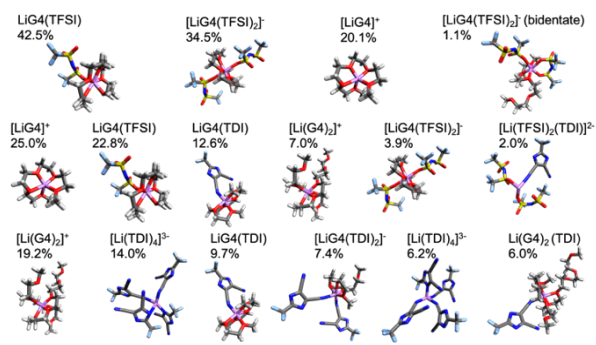

*Figure 5. Populations of local structures in LiTDI/LiTFSI in G4 at 1:1 salt:solvent molar ratio with 0% LiTDI (top row), 30% LiTDI (middle row) and 100% LiTDI (bottom row). Reprinted with permission from Batteries & Supercaps.[10]*

In addition to these two topological representations, CHAMPION also supports two simpler representations of the local structure of atoms or topologies that are familiar to most users. One is the coordination number (CN) of an atom type, which here is the average number of bonds of that atom type in the atomic bond graph (Fig. 6), and the other is the solvation number (SN), which here is the average number of bonds of a molecule in the molecular bond graph.

**Analysis #1 – Topology population fractions**

The probability of a given chemical species to be in a specific topology is computed as the average fraction of time in which each exemplar of the species is part of that topology. An upper bound on the standard error of this probability is obtained by counting each exemplar as a single sample in the error estimate

$$\delta x_i = \frac{\sigma_{x_i}}{\sqrt{n}} \qquad (5)$$

where $\delta x_i$ is the population fraction standard error, $\sigma_{x_i}$ is the sample standard deviation for the population fraction $x_i$, with the fraction of time steps of each exemplar spent in the given topology as the samples, and $n$ the number of exemplars.

If the trajectory analysed is long enough that each exemplar on average changes topology several times, this error bound is unnecessarily loose. In such cases, the correlation time for an exemplar remaining in the same topology can be used to estimate the statistical inefficiency so that each exemplar can provide several effective samples in Eq. (4), *e.g.* using the technique of block averaging.[31]

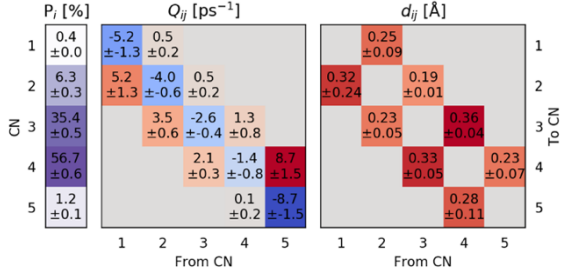

Figure 6. Properties related to Li+ CN for LiTFSI in ACN at 1:2 salt:solvent molar ratio: CN population (left), CN transition rate matrix (center), and mean distance travelled per transition between different CNs. Reprinted with permission from the Journal of the Electrochemical Society.[11]

**Analysis #2 – Bond and Topology lifetimes**

The mean lifetimes of different bond types and topologies are both estimated by fitting survival probability $P(\tau)$ as a function of time since birth $\tau$, sampled as the fraction of exemplars that survived a time equal or longer than $\tau$ out of all exemplars with a time of birth $< T - \tau$ for a trajectory of length $T$, to a stretched exponential function

$$\hat{P}(\tau) = \exp(-(\alpha\tau)^\beta) \qquad (6)$$

where $\hat{P}(\tau)$ is the fitted function with parameters α and β. Birth and death events are assumed to be independent and identically distributed (IID).
The mean lifetime according to the distribution of Eq. (6) is

$$\bar{\tau} = \frac{1}{\alpha\beta}\Gamma\left(\frac{1}{\beta}\right) \qquad (7)$$

The uncertainty in survival probability as a function of time is estimated as the product of the probability at the mean lifetime and the root-mean-squared relative error over the stretched exponential

$$\delta P^2 = \hat{P}(\bar{\tau})^2 \frac{1}{N}\sum_{n=1}^{N}\frac{\left(\hat{P}(\tau_n) - P(\tau_n)\right)^2}{P(\tau_n)^2} \qquad (8)$$

where $\delta P^2$ is the squared standard error of the probability and $n$ runs over the $N$ sampled lifetimes with $\tau_n$ being the $n^{\text{th}}$ observed lifetime and $P_n$ the sampled probability for the same lifetime. The uncertainty in mean lifetime is estimated by scaling $\delta P$ with the inverse derivative of $\hat{P}$:

$$\delta\bar{\tau} = \left(\frac{d\hat{P}}{d\tau}\right)^{-1}\delta\hat{P}(\bar{\tau}) \qquad (9)$$

**Analysis #3 – Topology transition rates**

The total transition rate away from a topology $i$ is simply the inverse of the average lifetime for that topology, computed according to the previous subsection:

$$\sum_{j\neq i} Q_{ji} = \frac{1}{\bar{\tau}_i} \qquad (10)$$

where $Q_{ji}$ is the rate of transitions from $i$ to $j$ and $j$ goes over all other topologies to which the species can transition. The uncertainty of the estimate for $Q_{ji}$ can be computed by standard error propagation for division from the uncertainty in $\bar{\tau}_i$

$$\delta\sum_{j\neq i} Q_{ji} = \frac{\delta\tau_i}{\bar{\tau}_i^2} \qquad (11)$$

The individual $Q_{ji}$ are proportional to the total number of transitions from $i$ to $j$ with the same constant of proportionality $C$ for all $j$

$$Q_{ji} = Cm_{ji} \qquad (12)$$

where $m_{ji}$ is the number of transition events and $C$ is uniquely determined by the requirement that the individual transition rates add up to the total outgoing transition rate from topology $i$. In an analogous manner, the squared standard errors of the individual transition rates from $i$ add up to the squared standard error of the total outgoing transition rates with each squared standard error proportional to the number of observed events, based on the assumption of IID transition events.

**Analysis #4 – Diffusivities of species**

The diffusivities of atoms, molecules and other species that do not change their topology, i.e. connectivity, during their trajectory are computed using the mean squared displacement (MSD)

$$\text{MSD}_X(\tau) = \langle\Delta\vec{r}^k(t, t+\tau)^2\rangle_{k,t} \qquad (13)$$

for species $X$, where

$$\Delta\vec{r}^k(t, t+\tau) = \int_t^{t+\tau}\vec{v}^k(t')dt \qquad (14)$$

is the displacement of exemplar $k$ of species $X$ between points in time $t$ and $t+\tau$, and the average runs over exemplars $k$ and time $t$. The MSD forms a curve that is expected to consist of an initial quadratic curve in the ballistic regime, followed by a straight line in the diffusive regime, past the diffusion onset time $\tau_D$. The diffusivity of $X$, $D_X$, is proportional to the slope of this curve. Most commonly, diffusivity is computed from the MSD using

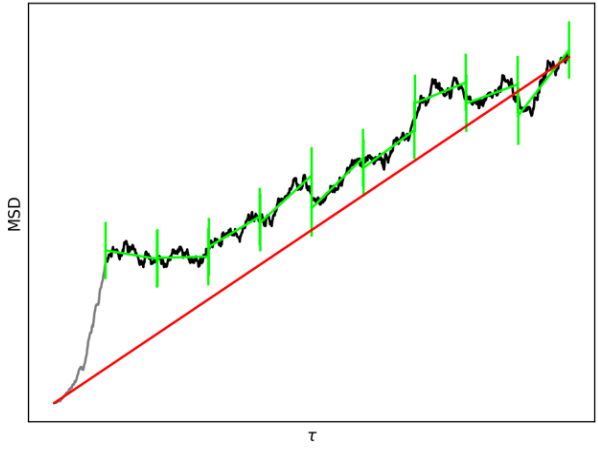

*Figure 7. Schematic comparison between the conventional and CHAMPION methods for assessing D from MSD. Conventionally (red) D is based on the MSD for the longest available interval. CHAMPION (green) instead first identifies the diffusive part of the MSD curve (black) and computes D based on averaging the slope of the MSD over the whole range of $\tau > \tau_D$.*

$$D_X = \lim_{\tau \to \infty} \frac{\text{MSD}_X}{6\tau} \quad (15)$$

which fails to exclude the ballistic part of the curve, and is also not a very robust approximation of the slope of the MSD, since it relies on a single finite difference, and it cannot be used to quantify uncertainty. CHAMPION instead bases its computation of the diffusivity of a species on the derivative of the MSD curve,

$$\frac{d\text{MSD}}{d\tau}(\tau) = \frac{d}{d\tau}\langle \Delta \vec{r}^k(t, t+\tau)^2 \rangle_{k,t} \quad (16)$$

The diffusivity is then evaluated as

$$D_X = \frac{1}{6}\left\langle \frac{d\text{MSD}}{d\tau} \right\rangle_{\tau > \tau_D} \quad (17)$$

where the average goes over all values of $\tau > \tau_D$. This derivative is evaluated for each sampled value of τ (Fig. 7). The mean of the samples is the estimated diffusivity and their standard error

$$\delta D_X = \frac{\sigma_{D_X}}{\sqrt{N}} \quad (18)$$

where $\sigma_{D_X}$ is the sample standard deviation and $N$ is the number of samples, after accounting for statistical uncertainty, assuming that the samples are IID, which is supported by the white noise appearance of this curve.

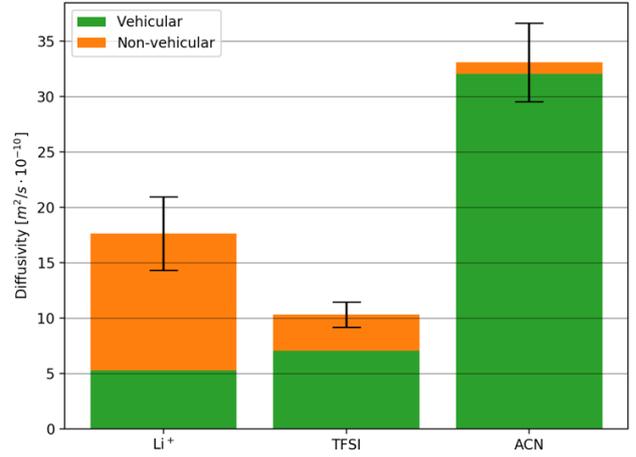

*Figure 8. Diffusivity of the different species in LiTFSI in ACN at 1:2 salt:solvent molar ratio, also decomposed into vehicular (green) and non-vehicular (orange) contributions (here rotational contributions were all negligible). Error bars refer to the total. Reprinted with permission from the Journal of the Electrochemical Society.[11]*

**Analysis #5 – Diffusivity contributions from different modes of motion**

The total diffusivity of species $X$ can be additively decomposed into contributions from different modes of motion: vehicular, rotational and structural (Fig. 8). Vehicular motion is the motion of atom or molecule $k$ due to translation of the centre-of-mass (CoM) of its current topology, i.e. its "vehicle", rotational motion is due to rigid body rotation of the topology about its CoM, and structural motion is the non-rigid body motion of $k$ within its topology.

To compute diffusivities due to different modes of motion, the instantaneous velocity is first decomposed,

$$\vec{v}^k(t) = \vec{v}^k_{\text{vehicular}}(t) + \vec{v}^k_{\text{rotational}}(t) + \vec{v}^k_{\text{structural}}(t) = \sum_m \vec{v}^k_m(t) \quad (19)$$

We then explicitly carry out the derivative of the MSD w.r.t. τ

$$\frac{d}{d\tau}\left\langle \left(\Delta \vec{r}^k(t, t+\tau)\right)^2 \right\rangle = 2\langle \Delta \vec{r}^k(t, t+\tau) \cdot \vec{v}^k(t+\tau) \rangle \quad (20)$$

where the averages go over $k$ and $t$ and the expression is a function of τ. We finally decompose the velocity and insert the result into Eq. (17):

$$D^*_{X,m} = \frac{1}{3}\langle \Delta \vec{r}^k(t, t+\tau) \cdot \vec{v}^k_m(t+\tau) \rangle_{k,t,\tau>\tau_D} \quad (21)$$

so that

$$D_X = \sum_m D^*_{X,m} \quad (22)$$

as intended. As for the total diffusivity of the species (see above) the averages with respect to $k$ and $t$ can be carried out first, and the samples as function of $\tau$ be used to assess the uncertainty assuming IID samples.

**Analysis #6 – Diffusivity contributions from different topologies**

A different way to decompose the diffusivity, which can be combined with the above analysis, is into contributions from different topologies, where the instantaneous velocity is decomposed by

$$\vec{v}^k(t) = \sum_i \chi_i^k(t)\vec{v}^k(t) \equiv \sum_i \vec{v}_i^k(t) \qquad (23)$$

into the contributions for each topology, $i$, where the indicator function $\chi_i^k(t)$ is 1 for the current topology and 0 for all others. In analogy with the previous analysis, the diffusivity contributions from the different topologies are given by

$$D^*_{X,i,\text{tot}} = \frac{1}{3}\langle \Delta \vec{r}^k(t, t+\tau) \cdot \vec{v}_i^k(t+\tau)\rangle \qquad (24)$$

The average should here in principle run over $k$, $t$ and $\tau$ as usual, but the computational complexity scales linearly with the number of topologies, making it potentially prohibitive to average over all variables, even more so if the diffusivity is decomposed along both modes of motion and topologies. A more pragmatic approach, utilizing that Eq. (24) is valid also for any specific value of $\tau > \tau_D$, is to choose a specific $\tau$ large enough to clearly be in the diffusive regime, but small enough that $\tau \ll T$, for trajectory length $T$, so that many uncorrelated values of $t$ may be sampled for the chosen $\tau$. This considerably reduces the computational cost while still sampling the entire trajectory. It becomes necessary, however, to account for the fact that velocities change with a characteristic timescale such that velocity at nearby points in time is correlated. We measure the statistical inefficiency of velocity using the methods of block averages to quantify the effective number of uncorrelated samples, and compute standard errors based on this number.[31]

**Analysis #7 – Diffusivity contributions from transitions between topologies**

Yet another way to decompose the diffusivity is into contributions from both topologies and the transitions between them. Above we described how to compute the total contribution of each topology, and this included also contributions more naturally attributable to transitions between topologies. The typical example would be a solvent exchange event. In this subsection we elaborate further on this and also how to distinguish between the vehicular, rotational and structural diffusivity contributions from different topologies and the transitions between them, i.e. diagonal and non-diagonal terms with respect to the topologies.

Starting with what to attribute to diagonal *vs.* non-diagonal contributions, ideally, in the case of stable topologies, in the long-time limit used to define diffusivity, the MSD of the topology (i.e. the vehicular displacement) tends to infinity linearly, whereas the expected rotational and structural displacements are constant and small at the long-time limit. The vehicular displacement therefore tends towards the total displacement in relative terms, whereas the rotational and structural parts eventually yield a negligible relative contribution. At the opposite extreme, consider *e.g.* ion transport by hopping between discrete coordination sites in a crystal lattice where the topology is immobilised, and it makes little sense to attribute the ion transport to the individual topologies with barely moving CoM. Thus, the transport is fully by transitions between topologies and completely non-diagonal. Another example is proton transport in aqueous media including that of polyelectrolytes such as Nafion-based membranes for fuel cells, which in part occurs vehicularly by hydronium ions, but also by the Grotthuss mechanism.[44] The latter consists of rotation of the hydronium ion followed by a proton hop to a water molecule to create a new hydronium ion. Hence, the overall transport is a combination of vehicular, rotational, and structural events, where the vehicular part is a diagonal contribution and the rotational and structural parts are non-diagonal.

For the vehicular contribution, we use the same indicator functions as described above, while for the non-diagonal contributions we use an indicator function $\xi_{ij}(t)$ for transitions from topology $j$ to topology $i$ starting from the midlife point of the original topology, and ending at the midlife point of the resulting topology. This is chosen as the least arbitrary break points, being as far as possible away from the transition times, but it is also the case that the exact break point becomes irrelevant if there is on average a large number of cycles between transition events, as the expected net displacement of a vibration or rotation cycle that does not lead to a transition is close to zero. Our final results for the analysis of diagonal and non-diagonal diffusivity contributions is

$$D^*_{X,i} = \frac{1}{3}\langle \Delta \vec{r}^k(t, t+\tau) \cdot \chi_i(t+\tau)\vec{v}^k_{\text{vehicular}}(t+\tau)\rangle \qquad (25)$$

and

$$D^*_{X,ij} = \frac{1}{3}\langle \Delta \vec{r}^k(t, t+\tau) \cdot \xi_{ij}(t+\tau)(\vec{v}^k_{\text{rotational}} + \vec{v}^k_{\text{structural}})(t+\tau)\rangle \qquad (26)$$

**Concluding Remarks**

We have here presented the design and methodology of our novel software CHAMPION, which allows analyses of atomic trajectories to be based on dynamic structure discovery. This unique approach gives a time-resolved view of the studied system at the intermolecular level where many technologically relevant properties arise. The combination of automated dynamic structure discovery to find both reversible and irreversible bonds, graph theory to classify all topologically distinct connected components and local structures, and statistical physics to characterise the properties of the discovered structures, has the potential to elucidate much of the complexity in dynamic and disordered materials and liquids. This is necessary in order to understand many dynamical processes such as transport, self-organisation, phase transitions etc. This should contribute to progress in fundamental understanding of many technologically and scientifically important condensed matter systems, and also aid in their rational design. Application and development in a much broader context than for the battery electrolyte studies for which it was developed are thus both expected and foreseen. CHAMPION will eventually be made more broadly available and further development plans include a graphical user interface (GUI), a database integration, and Python language bindings, which all should increase its ease-of-use.

## List of acronyms

| | |
|---|---|
| ACN | acetonitrile |
| AIMD | *ab initio* MD |
| CHAMPION | Chalmers Hierarchical Atomic, Molecular, Polymeric & Ionic Analysis Toolkit |
| CoM | Centre-of-Mass |
| DMC | dimethyl carbonate |
| DoF | Degree of Freedom |
| EC | ethylene carbonate |
| ESW | Electrochemical Stability Window |
| G4 | tetraglyme |
| GPE | Gel Polymer Electrolyte |
| HCE | Highly Concentrated Electrolyte |
| IID | Independent, Identically Distributed |
| IL | Ionic Liquid |
| LHCE | Localized HCE |
| LIB | Lithium-Ion Battery |
| LiTDI | lithium 2-trifluoromethyl-4,5 dicyanoimidazole |
| LiTFSI | lithium bis(trifluoromethanesulfonyl) imide |
| MD | Molecular Dynamics |
| MSD | Mean Squared Displacement |
| pRDF | partial RDF |
| RDF | Radial Distribution Function |
| SIL | Solvate Ionic Liquid |
| SPE | Solid Polymer Electrolyte |
| vdW | van der Waals |

## Conflicts of interest

RA, FÅ and PJ are co-founders and own shares in Compular AB.

## Acknowledgements

The research presented has received funding through the HELIS project (European Union's Horizon 2020 research and innovation program under Grant Agreement No. 666221) and the Swedish Energy Agency grants (#P43525-1 and #P39909-1). P. J. would also like to acknowledge several of Chalmers Areas of Advance: Materials Science, Energy, and Transport, for continuous support, and specifically the Theory and Modelling scheme of Advanced User Support to R.A. A.A.F. acknowledges the Institut Universitaire de France for the support.